\documentclass[preprint,aps,onecolumn,floatfix,showpacs]{revtex4}
\usepackage{graphicx}
\begin{document}

\title{Helicity-dependent photoabsorption cross sections on the nucleon
} 

\author{
R.A.~Arndt, W.J.~Briscoe, I.I.~Strakovsky, and R.L.~Workman
}

\address{Center for Nuclear Studies, Physics Department \\
The George Washington University, Washington, D.C. 20052-0001
}

\draft
\date{\today}

\begin{abstract}

We examine the energy dependence of single-meson photoproduction as
it contributes to the Gerasimov-Drell-Hearn (GDH) sum rule. For photon 
energies above approximately 1~GeV, through the full resonance region, 
this contribution dominates the proton sum rule integral. Over the same
energy region, our single-pion contribution to the neutron sum 
rule also qualitatively follows a recent set of GDH data. 
The predicted neutral-pion contribution is nearly zero above 1~GeV in 
this result. The SAID and Mainz (MAID) results are very different for 
a number of observables over this energy region.

\end{abstract}

\pacs{13.60.Hb, 25.20.Lj}

\maketitle

\section{Introduction}

The Gerasimov-Drell-Hearn (GDH) sum rule~\cite{gdh} states that
\begin{eqnarray}
{{2\pi^2 \alpha}\over {M^2}}\left( \kappa_{p,n} \right)^2 
= \int^{\infty}_{\nu_0} {{\sigma_{3/2}(\nu ) - \sigma_{1/2}(\nu )}\over 
{\nu}} d\nu ,
\label{1}\end{eqnarray}
where $\kappa_p ( \kappa_n )$ is the proton(neutron) anomalous magnetic moment,
$\nu$ is the laboratory photon energy, $M$ is the nucleon mass, and $\alpha$ is 
the fine structure constant. The left-hand-side represents the single-nucleon 
contribution to the spin-flip amplitude, while the right-hand-side involves an 
integration over the difference of helicity 3/2 and 1/2 $\gamma$-nucleon total 
cross sections. This relation can be derived on the basis of fundamental 
principles (such as gauge and Lorentz invariance) and is expected to be 
satisfied exactly.  While early results~\cite{old} suggested a possible 
violation, careful measurements and estimates of the important high-energy 
contribution have greatly reduced this discrepancy~\cite{Dutz2}. It now seems 
unlikely that the sum rule, and the fundamental assumptions used in its 
derivation, are violated. 

\begin{figure*}[th]
\centering{
\includegraphics[height=0.7\textwidth, angle=90]{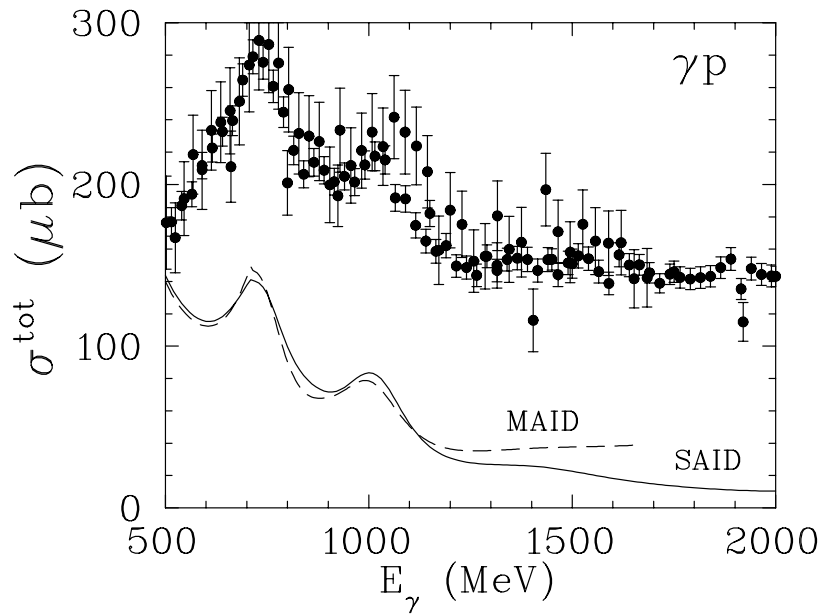}
}\caption{Single-pion photoproduction contributions to the
          total cross section for a proton target.  SAID
          SM05~\protect\cite{SAID05} (solid) and MAID2003
          ~\protect\cite{MAID} (dashed) analyses.  The proton
          photoabsorption data from Ref.~\protect\cite{pdgd}.}  
          \label{fig:g1}
\end{figure*}

\begin{figure*}[th]
\centering{
\includegraphics[height=0.5\textwidth, angle=90]{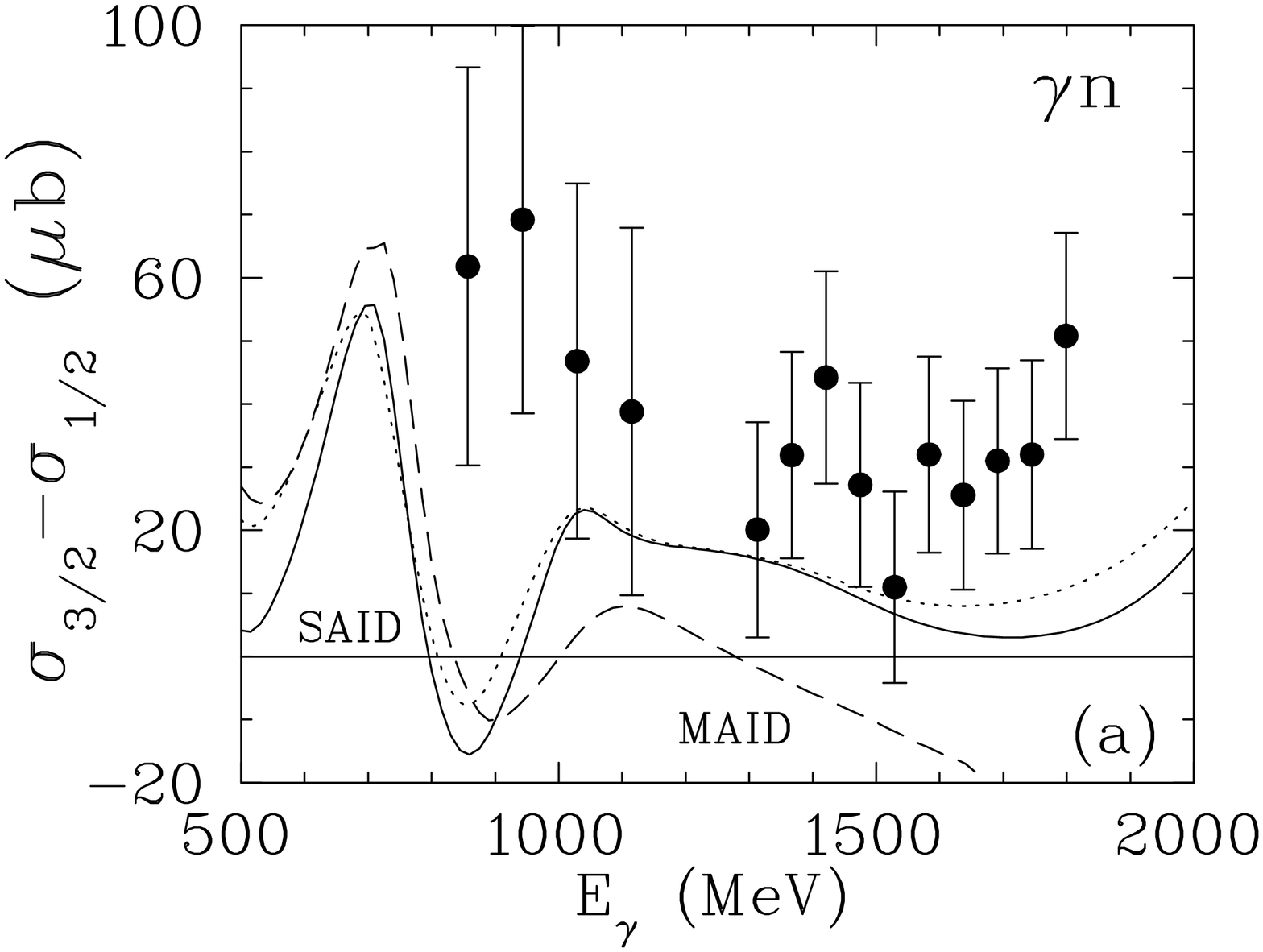}\hfill
\includegraphics[height=0.5\textwidth, angle=90]{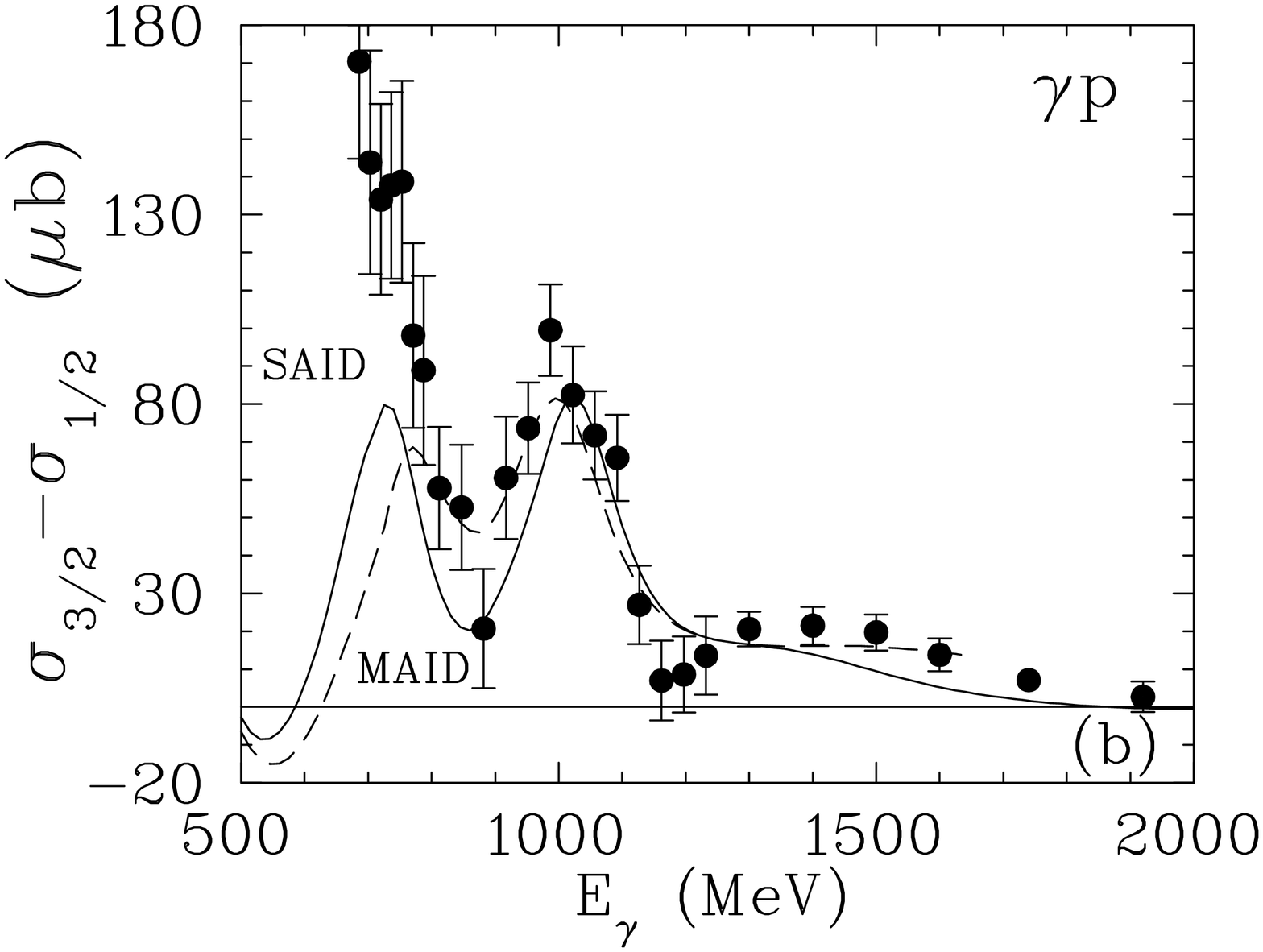}
}\caption{Single-pion photoproduction contributions to the
          (a) neutron and (b) proton GDH sum rules from the
          SAID SM05~\protect\cite{SAID05} (solid),
          recently published SM02~\protect\cite{SAID} (dotted), 
          and MAID2003~\protect\cite{MAID} (dashed) analyses.  
          GDH data from Refs.~\protect\cite{Dutz,Dutz2}.} 
          \label{fig:g2}
\end{figure*}

\begin{figure*}[th]
\centering{
\includegraphics[height=0.5\textwidth, angle=90]{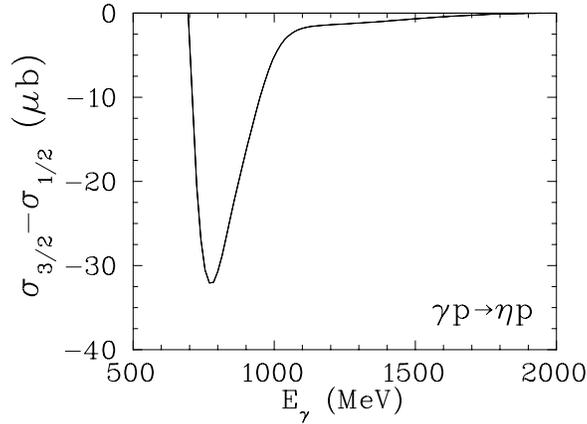}\hfill
}\caption{SAID single-eta photoproduction contribution to the proton
          sum rule.}
          \label{fig:g3}
\end{figure*}

Early estimates of the GDH integral used phenomenological single-pion 
photoproduction amplitudes and crude estimates for two-pion and other-meson 
production. Particularly important are the two-pion production contributions, 
as they dominate the total photoabsorption cross section over much of the 
resonance region. To illustrate this point, the SAID~\cite{SAID} and 
MAID~\cite{MAID} single-pion contributions to the total photoabsorption cross 
section have been plotted in Fig.~\ref{fig:g1}.  Some of these two-pion 
channels have been measured separately, and would appear to give a significant 
contribution to the GDH sum rule over much of the resonance region~\cite{2pi}. 
However, a quite different conclusion results if the measured helicity-dependent 
total cross section ($\sigma_{3/2} - \sigma_{1/2}$) is compared to its 
single-pion contribution over the resonance region. Single-pion predictions for 
the SAID and MAID analyses are again plotted for comparison in Fig.~\ref{fig:g2}.
The contribution from eta-meson photoproduction is significant near its
threshold, due to the dominant N(1535), but is expected to contribute 
little at higher energies. Its contribution, obtained from a SAID fit to
eta-photoproduction data, is shown in Fig.~\ref{fig:g3}. The contributions from 
kaon photoproduction are also small, based on fits to the existing data.

\begin{figure*}[th]
\centering{
\includegraphics[height=0.53\textwidth, angle=90]{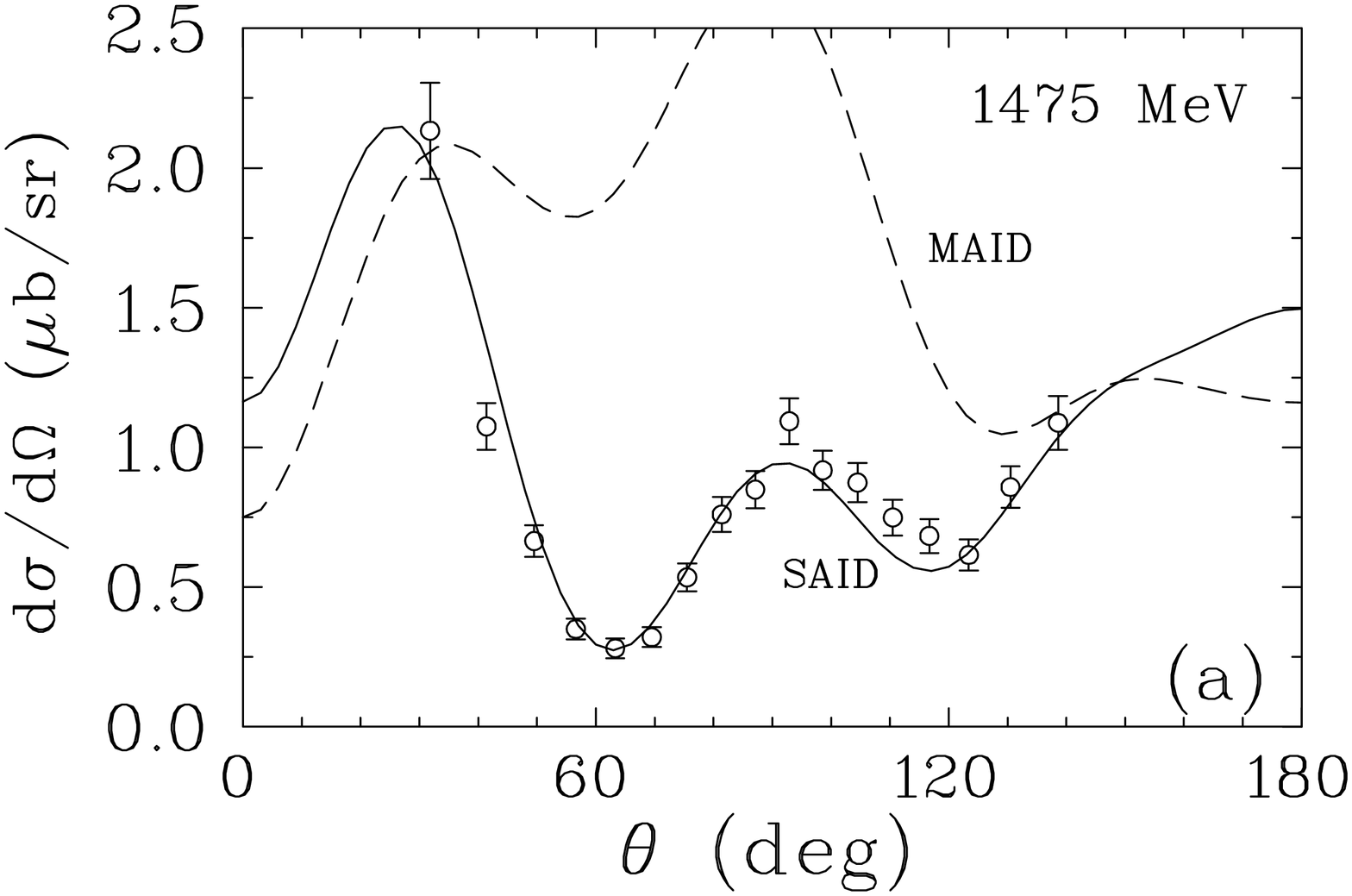}\hfill
\includegraphics[height=0.47\textwidth, angle=90]{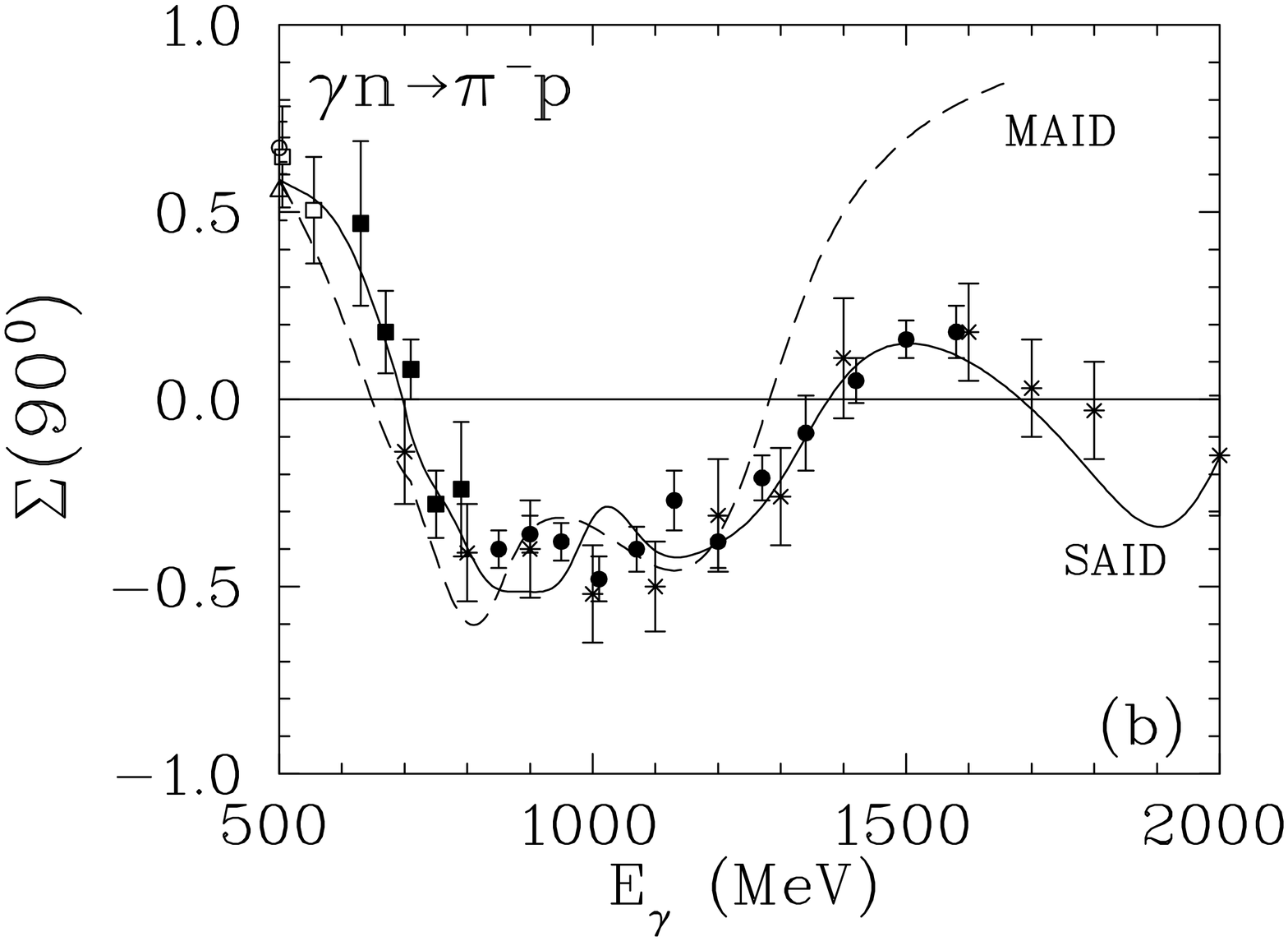}
}\caption{Single-pion photoproduction above 1~GeV.
          (a) $p\pi^0$ differential cross section at
          1475~MeV and (b) $p\pi^-$ $\Sigma$ beam
          asymetry at 90$^\circ$.  Cross section data
          from ~\protect\cite{ELSA} and  $\Sigma$ data
          from~\protect\cite{s90}.  Predictions from
          SAID SM05~\protect\cite{SAID05} (solid) 
          and MAID2003~\protect\cite{MAID} (dashed) 
          analyses.} \label{fig:g4}
\end{figure*}

In the proton sum rule, both the SAID and MAID single-pion predictions account 
for almost the entire result for photon energies above 1~GeV. The SAID results 
extend over a larger energy region, but qualitatively agree with MAID where 
comparisons are possible.  The SAID single-pion contribution to the neutron 
sum rule also tends to follow the GDH measurement. Here, however, the MAID fit 
diverges from the data, which (assuming a correct single-pion part) would 
imply a large and increasing two-pion contribution to the neutron.  The dip 
near 900~MeV is seen in both the SAID and MAID single-pion contributions to 
the neutron sum rule and does not appear to be the result of a problem in the 
multipole analysis. However,  
other quantities predicted by the SAID and MAID fits are 
often quite different, particularly above 1~GeV in the photon energy. In 
Figs.~\ref{fig:g4}(a) and \ref{fig:g4}(b), we compare the SAID and MAID fits 
to $p\pi^0$ differential cross section and $p\pi^-$ $\Sigma$ data. 

\begin{figure*}[th]
\centering{
\includegraphics[height=0.5\textwidth, angle=90]{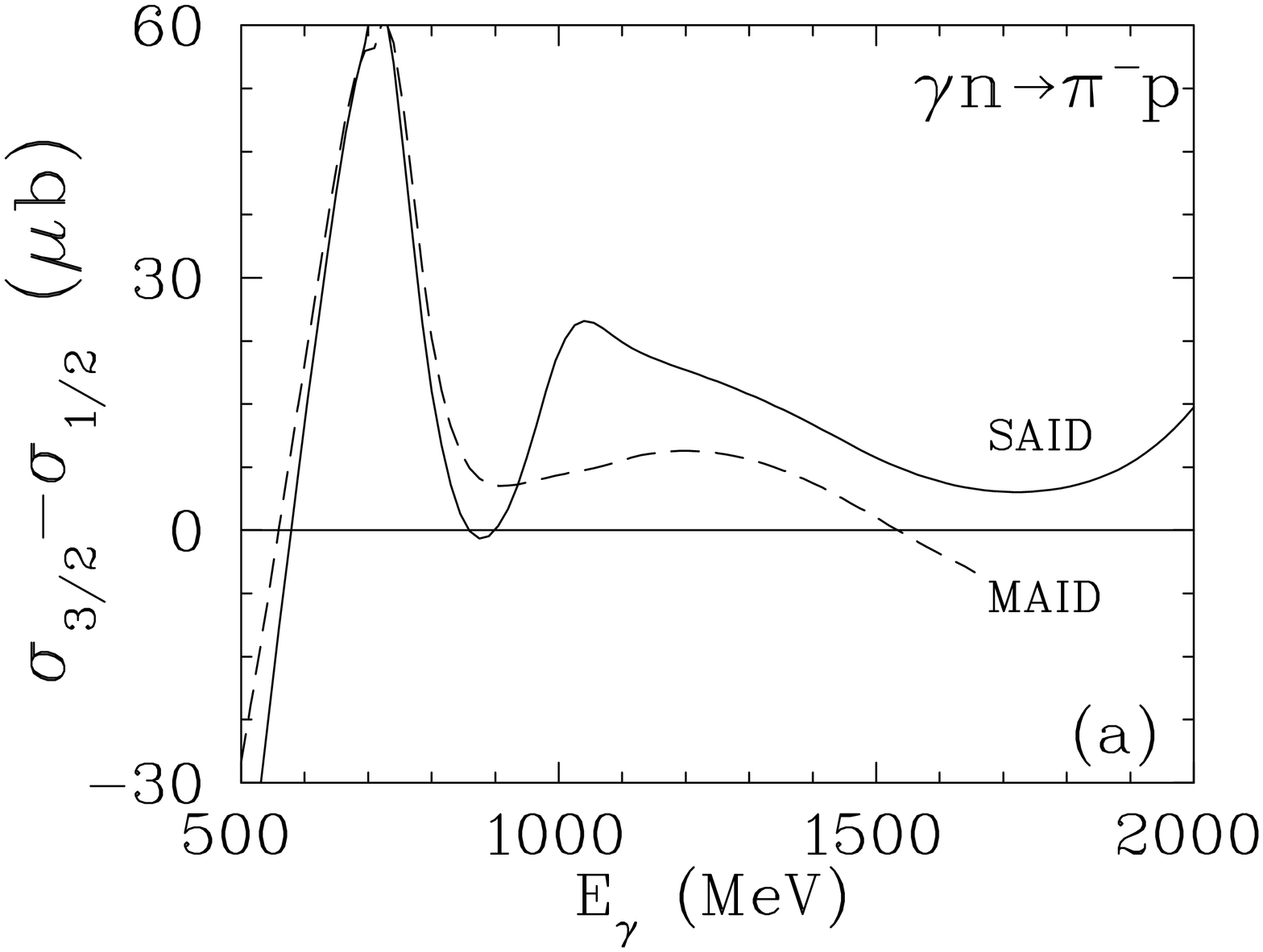}\hfill
\includegraphics[height=0.5\textwidth, angle=90]{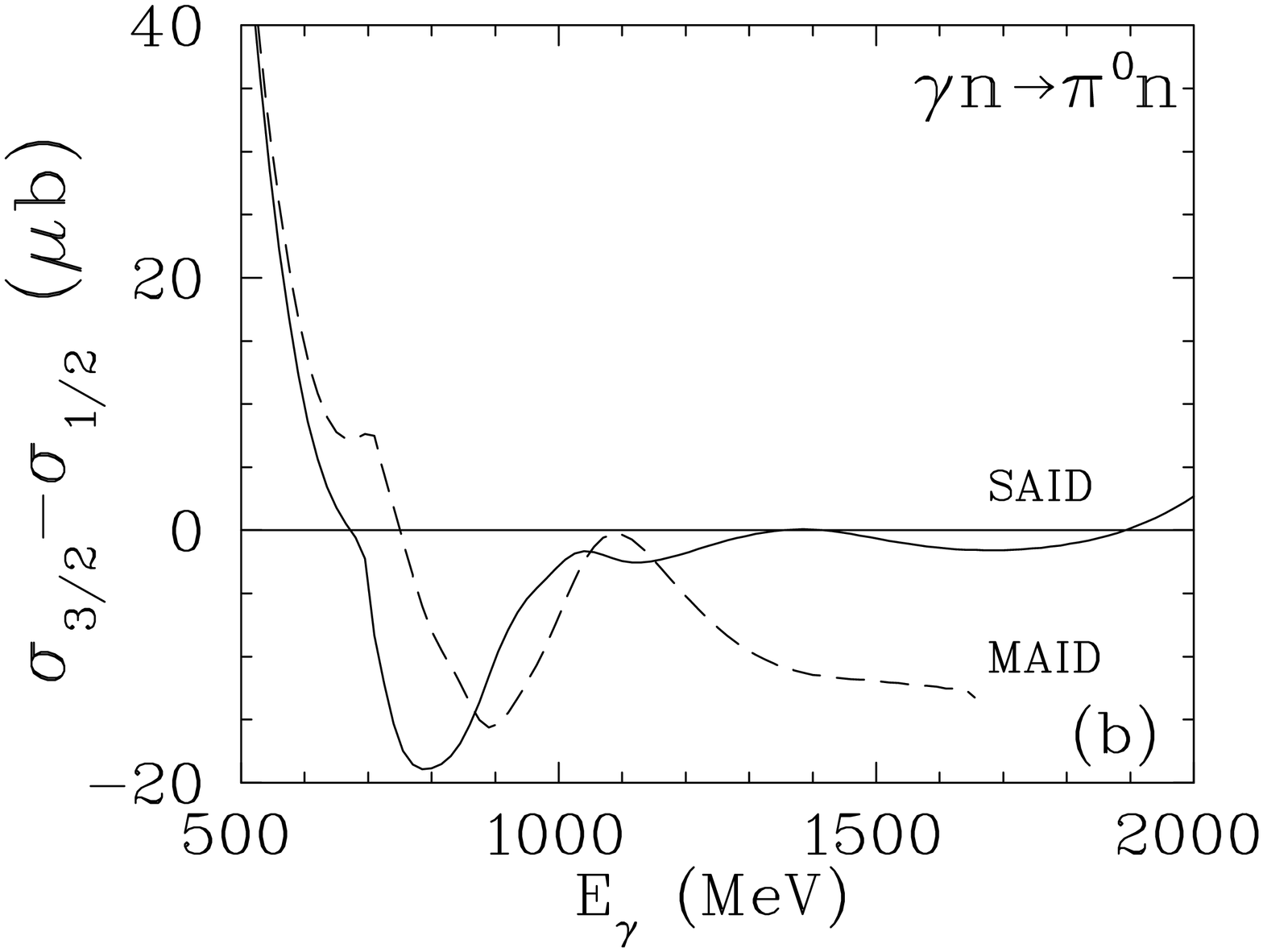}
}\caption{Single-pion photoproduction for (a) $\gamma n\to\pi^-p$ 
          and (b) $\gamma n\to\pi^0n$.  Predictions are given
          for the SAID SM05~\protect\cite{SAID05} (solid) 
          and MAID2003~\protect\cite{MAID} (dashed) analyses.}  
          \label{fig:g5}
\end{figure*}

Conclusions regarding single-pion contributions to the neutron sum rule 
are much less solid, as the underlying single-pion photoproduction database
is less complete and precise than the available proton-target data. In 
particular, there is almost no $\gamma n\to n\pi^0$ data of any kind. This 
component of the neutron sum rule is therefore purely a prediction, based 
upon amplitudes extracted from the $p\pi^0$, $n\pi^+$, and $p\pi^-$ channels.  
Interestingly, these amplitudes conspire to cancel in the SAID solution, 
giving a negligible $n\pi^0$ contribution to the neutron sum rule above about 
1~GeV. The $n\pi^0$ and $p\pi^-$ contributions from SAID and MAID are compared 
in Fig.~\ref{fig:g5}. We therefore find that the $p\pi^-$ component mainly 
responsible for the qualitative agreement shown in Fig.~\ref{fig:g2}(a). Both 
the SAID solution and data suggest an upward trend near 2~GeV. Examining the 
behavior of our fits at the high energy limit, we cannot claim this to be 
more than an artifact. Comparing our fits SM02 and SM05, the fit influenced 
by more recent high-energy ELSA and JLab data (SM05) displays a less 
pronounced upward trend.  If the highest energy measurement is more than a 
fluctuation, it would signal a resonance contribution not contained in either 
the SAID pion-nucleon~\cite{pin} or pion-photoproduction~\cite{SAID} analyses. 
Given that a similar trend is not seen in the proton GDH data, this
resonance contribution would decay to $\gamma n$ much more readily than
$\gamma p$.

In summary, data that have been measured to verify the GDH sum rule may
be equally valuable in understanding both resonance physics and the
transition between resonance and Regge dominated regions. The apparent
cancellation of large multi-pion contributions to the proton GDH sum rule 
suggests an underlying principle which deserves further investigation.
Certainly it would be interesting to extend the single-pion analysis in 
order to see how high in energy the trend of Fig.~\ref{fig:g2} continues.  
Work on this project is in progress.  The comparison of proton and neutron 
GDH data, versus the single-pion photoproduction contribution, suggests the 
existence of states coupling more strongly to $\gamma n$ than $\gamma p$. 
These features should motivate further measurements of two-pion production, 
and single-meson photoproduction off the neutron. 

\begin{acknowledgments}
We thank P.~Pedroni for a discussion of the GDH measurements.
This work was supported in part by a U.S.~Department of Energy grant
DE--FG02--99ER41110, and by Jefferson Lab.
\end{acknowledgments}



\end{document}